\documentstyle[rotate,psfig,twocolumn]{mn}

\newcommand{\lapp}{\mbox{\raisebox{-0.3em}{$\stackrel{\textstyle <}{\sim}$}}}
\newcommand{\gapp}{\mbox{\raisebox{-0.3em}{$\stackrel{\textstyle >}{\sim}$}}}

\title[Radio sources at low latitudes]{Radio sources at low Galactic latitudes}

\author[D.J. Saikia et al.]{D.J. Saikia$^{1,2}$, P. Thomasson$^{2}$,
S. Roy$^{1}$, A. Pedlar$^{2}$ and T.W.B. Muxlow$^{2}$ \\
$^{1}$ Tata Institute of Fundamental Research, National Centre for Radio
Astrophysics, P.B. No. 3, Ganeshkhind, Pune 411 007, India \\
$^{2}$ Jodrell Bank Observatory, University of Manchester, Macclesfield, Cheshire, SK11 9DL, UK \\
}

\date{Received:}
\begin{document}
\maketitle

\begin{abstract}
High-resolution radio observations of a sample of 65 radio sources at low Galactic latitudes are presented.
The sources were all observed at 5 GHz with the VLA A-array. MERLIN observations at 5 GHz of the ultracompact 
\hbox{H\,{\sc ii}} region G34.26$+$0.15 and one of the extragalactic sources, B1857$-$000, are also presented, as
are GMRT observations of \hbox{H\,{\sc i}} in the direction of three sources, B1801$-$203, B1802$-$196 and B1938+229.
These observations were made with the objectives of (i) finding compact 
components suitable for studying the effects of interstellar scattering at lower frequencies, 
(ii) identifying high surface-brightness lobes of background radio sources to probe the Galactic magnetic 
field on different scales via polarization observations, and (iii) searching for young supernova remnants.  
The nature of the sources found to have shell or shell-like structure and exhibiting both thermal and 
non-thermal spectra are discussed. Of the remaining sources, B1749$-$281 is coincident within the positional errors 
of a known pulsar, not detected earlier at 5 GHz. The rest are likely to be background extragalactic objects. 
\end{abstract}


\section{Introduction}
In recent years, there have been a number of relatively high-resolution surveys of the Galactic plane
with angular resolutions of $\sim$5 arcsec at cm wavelengths (e.g. Garwood et al. 1988; Zoonematkermani
et al. 1990, hereinafter referred to as Z90; Helfand et al. 1992; Becker et al. 1994, hereinafter referred
to as B94). These surveys have led to the identification of new
ultracompact \hbox{H\,{\sc ii}} regions, planetary nebulae and supernova remnant candidates (cf. B94). 
Higher resolution observations of samples selected from these surveys could be useful
for identifying sources suitable for studying the ionized component and magnetic field in our Galaxy, 
and also for finding young supernova remnants.

The scattering of radio wave propagation
by turbulent fluctuations of electron density in the interstellar medium (ISM), referred to as interstellar 
scattering (ISS), has proved to be a useful technique in the study of the ionized component of the ISM. 
ISS and its effects have been reviewed extensively by Rickett (1990) and more recently discussed by Armstrong, Rickett
\& Spangler (1995). One of the observable effects of ISS is the apparent broadening of the angular diameter
of an intrinsically compact radio source, with the shape and size of the broadening being determined by the
nature and distribution of the turbulence. A number of heavily scattered sources such as Sgr A$^\ast$, 
Cygnus X-3, NGC6334B and others seen through the Cygnus region have been studied in some detail. There
has also been evidence of anisotropic turbulence inferred from measurements of very elliptical scattering disks
along many lines of sight through the ISM (e.g. Wilkinson, Narayan \& Spencer 1994; Rickett, Lyne \& Gupta 1997;
Spangler \& Cordes 1998; Trotter, Moran \& Rodri\'{g}uez 1998; Desai \& Fey 2001).

To enlarge upon such studies a programme has been started to make sub-arcsec resolution images of low-latitude
sources, primarily to find compact components suitable for studying the effects of ISS. It is worth noting that previous
studies of extraglactic sources have largely concentrated on flat-spectrum, core-dominated radio sources. These 
are likely to be inclined at small angles to the line of sight and hence have prominent nuclear jets. In such 
cases, source structure is likely to significantly complicate the analysis of the effects of ISS. There might be
less of a problem in lobe-dominated sources with cores which may not have prominent jets, but are strong enough 
for observations with telescopes such as the Very Large Array (VLA), the Multi-Element Radio Linked Interferometer
Network (MERLIN) and the Very Long Baseline Array (VLBA).

The other principal objectives of the work are to find young supernova remnants (e.g. Green 1985, 1989; 
Cowan et al. 1989; Sramek et al. 1992, and
references therein; Clark, Steele \& Langer  2000) and identify high surface-brightness lobes of 
background radio sources which could be used to probe the Galactic
magnetic field on different scales (e.g. Simard-Normandin \& Kronberg 1980; Sofue \& Fujimoto 1983; Lyne \& Smith 
1989; Clegg et al. 1992; Rand \& Lyne 1994). The higher angular resolution of our observations should enable a search
for younger, and less evolved supernova remnants (SNRs) than has been carried out in earlier surveys (Green 1989 and 
references therein, Sramek et al. 1992). Polarization observations of the high-brightness sources will add to the
existing samples (Roy, Rao \& Subrahmanyan 2004) which can be used to probe
small-scale turbulence and structure of the magnetic field, especially towards the inner Galaxy.

\section{Radio observations}
In this paper the results of observations of a sample of 65 sources with 0$^\circ$$\lapp$ {\it l} $\lapp$
60$^\circ$ and Galactic latitude $\lapp \left|3\right|$ $^\circ$ are presented. These sources were chosen from the catalogues of 
Clark \& Crawford
(1974) and Garwood et al. (1988), and a sample of sources observed by Anantharamaiah \& Narayan (1988 and private communication). 
The choice of the sources observed depended on the lack of structural information with sub-arcsec resolution, 
the brightness of the source (they were the brighter ones) and scheduling constraints.

\subsection{VLA observations} Each source was observed for
approximately 5 to 10 minutes with the VLA A-array at 5 GHz on 1989 January 31. The sources were observed in the
standard way with a phase calibrator being observed before and after the scans on two or three nearby sources.
The primary flux density calibrator was 3C286. The flux densities are in the scale of Baars et al. (1977).
All the data analyses were done using the Astronomical Image Processing Software (AIPS) of
the National Radio Astronomy Observatory.

\subsection{MERLIN observations} 
The ultracompact \hbox{H\,{\sc ii}} region, B1850$+$011 (G34.26+0.15), was observed with MERLIN  during the initial phase of
its upgrade on 1992 November 9
and 1992 November 12 with a single polarisation and a bandwidth of 10 MHz. Each observation was for 
approximately 11 hours, but this time was divided between the target and B1904+013, which was
the phase reference source. The image presented here has been made by combining the MERLIN data with 
archival VLA observations made on 1986 April 27 at 4885 MHz. 

Another source in the sample, B1857$-$000, which has a resolved core in the VLA image, was observed with 
MERLIN at 5 GHz on 2001 September 09 for $\sim$10 hours.  Data analysis was carried out using the Jodrell
Bank D-program and AIPS and the flux density calibrator was 3C286. 
The amplitude calibration was established
from observations of OQ208 and 3C286, 
the flux density of the latter being assumed to be 7.086 Jy on the shortest 
MERLIN baseline, which is equivalent to a total flux density of 7.381 Jy on the VLA scale. The flux density of 
OQ208 was determined to be 2.39 Jy.

\subsection{GMRT observations}
Observations of \hbox{H\,{\sc i}} absorption lines towards the three sources, B1801$-$203, B1802$-$196 and B1938+229 have been  
carried out with the Giant Metrewave Radio Telescope (GMRT) using the 
default spectral line mode of the correlator with 128 frequency channels and a bandwidth of 2 MHz
(velocity coverage $\pm$200 km s$^{-1}$). The radio source B1802$-$196 was observed on 2000 October 08, while B1801$-$203
and B1938+229 were observed on 2001 December 03.
Further details about the array can be found at the GMRT website at {\tt http://www.gmrt.ncra.tifr.res.in}.
For these observations 3C286 was observed as the primary flux density
calibrator and 3C287 as the bandpass calibrator. 3C287 was chosen since 
it is at high galactic latitude and the effects of Galactic \hbox{H\,{\sc i}}  absorption
on its spectrum are known to be less than 1 \% (Dickey, Terzian \& Salpeter 1978). 
Automatic system temperature measurements were not implemented, and the maps have been scaled by 
the ratio of the target source flux density
as observed by the NRAO VLA Sky Survey (NVSS) (Condon et al. 1998) to that estimated from our data. The error in the
flux density is $\sim$10 per cent.

Before making the channel maps, the AIPS task UVLSF was used to subtract a constant term from the {\it uv}-data
across the frequency channels, corresponding to the continuum emission. The GMRT has an
FX correlator and therefore `Gibbs ringing' due to any sharp feature in the spectrum dies away quickly.
Consequently, it was not necessary to apply spectral smoothening to the data. As variations in the observed line frequency
due to the earth's motion during the observing period were significantly less than a channel width, 
no diurnal Doppler corrections have been applied to the data.

\begin{figure}
\vbox{
  \psfig{file=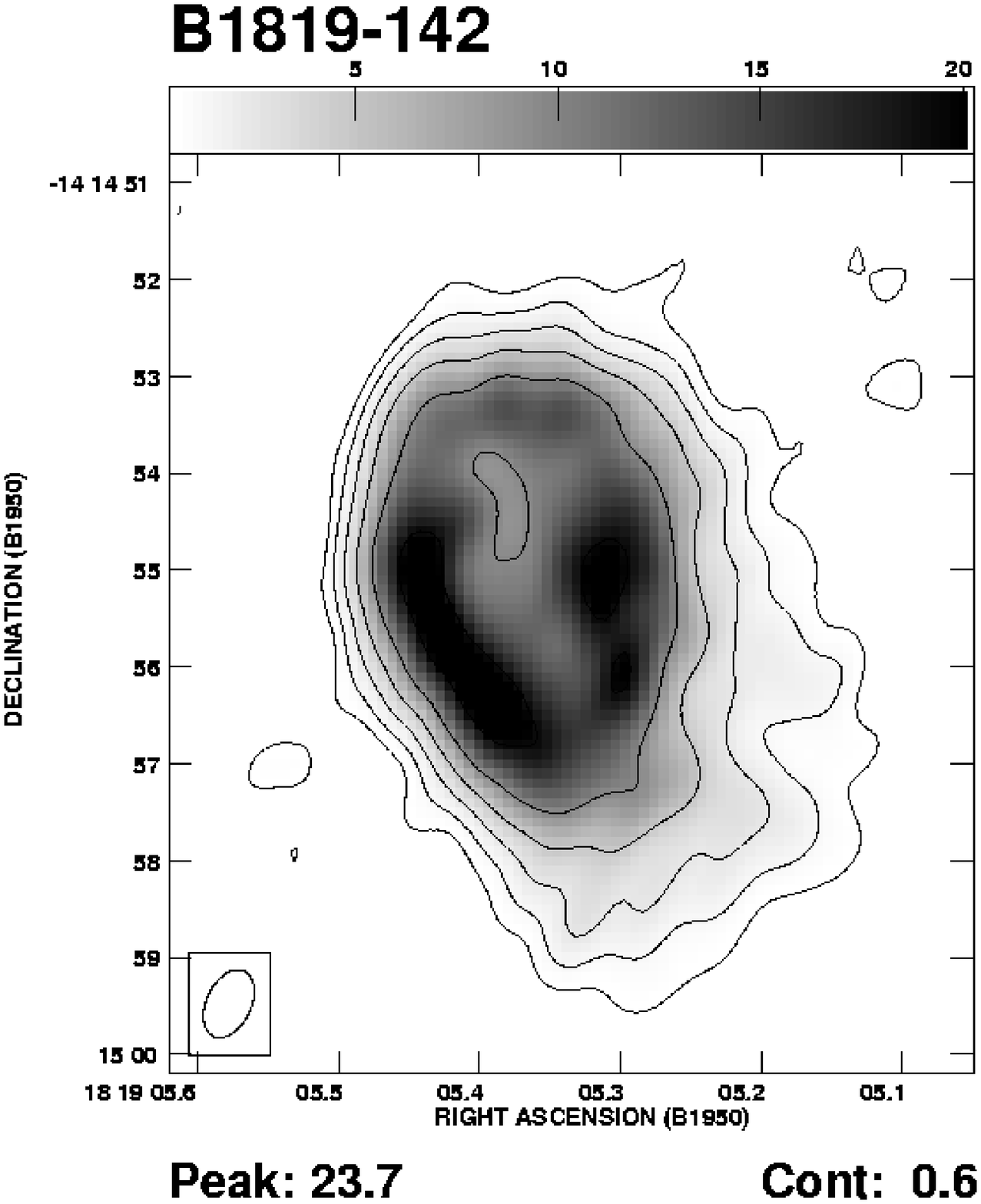,height=3.0in}
  \psfig{file=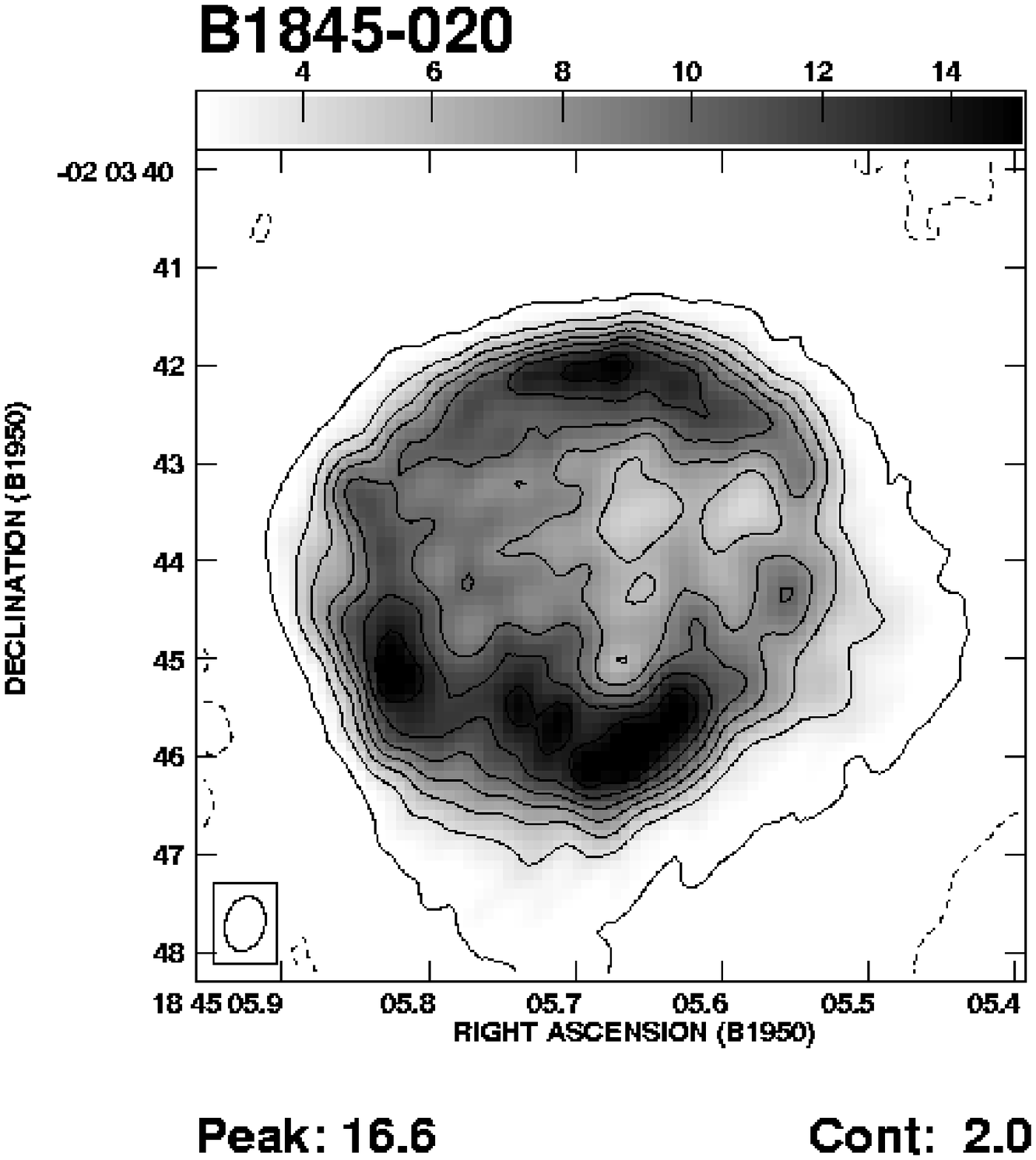,height=3.0in}
\caption[]{VLA images at $\lambda$6 cm of the ultracompact \hbox{H\,{\sc ii}} region B1819$-$142 (G16.94$-$0.07)
and a possible planetary nebula B1845$-$020 (G30.71$-$0.08).  The contours are $-$1, 1, 2, 4, 8 \ldots times
the first contour for B1819$-$142, and $-$1, 1, 2, 3, 4 \ldots times the first contour for B1845$-$020. 
The peak brightness and the level of the first contour in units of mJy/beam are shown below each image.}
}
\end{figure}

\section{Observational Results}

Some of the VLA observational parameters and observed properties of individual sources are presented in Table 1 which
is arranged as follows. Column 1: the source name using the B1950 co-ordinates. 
Columns 2 to 4: the half power beamwidth and orientation of the
restoring elliptical Gaussian beam. The major and minor axes are in arcsec and the position angle (PA) is in degrees.
Column 5: The total flux density of the source estimated by either setting a box around the source or summing the
flux densities of the individual components. Column 6: the component designation. 
Columns 7 to 12: the corresponding 
right ascension and declination of the radio intensity peaks in B1950 co-ordinates. Columns 13 and 14: peak
brightness and total flux density of the components in units of mJy beam$^{-1}$ and mJy respectively. The values for 
the components with a superscript $g$ in  Column 6 have been estimated from two-dimensional Gaussian fits. 
The remainder have been estimated by specifying boxes around the components. Column 15: classification of the
source. EG and G denote extragalactic and Galactic sources, and the letters U, SR, D and T denote an
unresolved, slightly resolved, double and triple sources respectively. A source is classified as a triple if the
central component appears to be the nucleus of a radio galaxy or quasar. C$+$E denotes a flat-spectrum core with
extended emission either in the form of a lobe or jet. For a flat-spectrum source, the spectral index, $\alpha$,
defined as S$\propto\nu^{-\alpha}$, is $\lapp$0.5.  A possible head-tailed radio source is indicated by the 
letters HT. Among Galactic objects, PSR denotes a pulsar, 
\hbox{H\,{\sc ii}} and PN denote an \hbox{H\,{\sc ii}} region and a planetary
nebula respectively. Column 16: An asterisk in this column indicates that there is a note to the source in 
Section 3 of the paper.

\begin{table*}
\caption{Observed properties of the sample of sources}
\begin{tabular}{l rrr r l rrr rrr  rr l l }
\hline
Source & \multicolumn{3}{c}{Beam size} & S$_t$ & Cmp. & \multicolumn{3}{c}{RA(B1950)} & \multicolumn{3}{c}{Dec.(B1950)} & S$_p^c$ & S$_t^c$ & Class.& Notes  \\
            &  maj    &  min    &   PA    &      &   & & & & & & & & & & \\
        & $^{\prime\prime}$ & $^{\prime\prime}$ & $^{\circ}$ & mJy & & h & m & s & $^{\circ}$ & $^{\prime}$ & $^{\prime\prime}$ & mJy/b & mJy & & \\
\hline

B1740$-$261  &  0.75   &  0.37   &   10    &     43     &   N$^g$  & 17 & 40 & 44.71 & $-$26 & 09 & 41.0 &     18    &      38   &  EG-D?   &           \\
            &         &         &         &            &   S$^g$  &    &    & 44.72 &       &    & 42.6 &    3.4    &     5.0   &          &           \\
B1748$-$253  &  0.62   &  0.36   &  161    &    492     &   C$^g$  & 17 & 48 & 45.78 & $-$25 & 23 & 17.5 &    413    &     492   &  EG-SR   &  $\ast$   \\ 
B1749$-$224  &  0.70   &  0.37   &    8    &    383     &   W      & 17 & 49 & 31.36 & $-$22 & 29 & 34.8 &     34    &     171   &  EG-D    &           \\
            &         &         &         &            &   E      &    &    & 32.49 &       &    & 31.8 &     23    &     163   &          &           \\
B1749$-$281  &  0.79   &  0.35   &    8    &    6.3     &   RS$^g$ & 17 & 49 & 49.24 & $-$28 & 06 & 00.1 &    5.6    &     6.3   &  G-PSR   &  $\ast$   \\
B1752$-$235  &  0.81   &  0.40   &   11    &    139     &   W      & 17 & 52 & 45.38 & $-$23 & 32 & 57.0 &    4.3    &      75   &  EG-T    &           \\
            &         &         &         &            &   C      &    &    & 45.69 &       &    & 56.1 &    6.0    &           &          &           \\
            &         &         &         &            &   E      &    &    & 45.91 &       &    & 55.5 &    4.5    &      48   &          &           \\
B1754$-$226  &  0.82   &  0.36   &  164    &     13     &   C$^g$  & 17 & 54 & 26.10 & $-$22 & 38 & 44.4 &     12    &      13   &  EG-U    &  $\ast$   \\
B1756$-$231  &  0.97   &  0.38   &  155    &     77     &   N      & 17 & 56 & 45.43 & $-$23 & 10 & 30.5 &    2.0    &      13   &  EG-D    &           \\
            &         &         &         &            &   S      &    &    & 45.39 &       &    & 36.2 &     38    &      64   &          &           \\
B1801$-$180  &  0.86   &  0.45   &  151    &     67     &   W      & 18 & 01 & 34.39 & $-$18 & 00 & 49.5 &    8.2    &      24   &  EG-D    &           \\
            &         &         &         &            &   E      &    &    & 35.07 &       &    & 42.6 &     13    &      43   &          &           \\
B1801$-$203  &  0.51   &  0.37   &  152    &    104     &   W      & 18 & 01 & 17.85 & $-$20 & 20 & 54.9 &     16    &      53   &  EG?-D   &  $\ast$   \\
            &         &         &         &            &   E      &    &    & 18.24 &       &    & 50.9 &    2.2    &      37   &          &           \\
B1802$-$196  &  0.75   &  0.39   &   11    &    249     &   W      & 18 & 02 & 08.83 & $-$19 & 36 & 21.5 &     11    &     181   &  EG-D    &  $\ast$   \\
            &         &         &         &            &   E      &    &    & 09.62 &       &    & 13.5 &    2.6    &      73   &          &           \\
B1804$-$141  &  0.65   &  0.38   &   10    &     68     &   W      & 18 & 04 & 55.17 & $-$14 & 06 & 52.9 &    1.2    &      35   &  EG-D    &           \\
            &         &         &         &            &   E      &    &    & 56.04 &       & 07 & 00.5 &    1.8    &      37   &          &           \\
B1808$-$184  &  0.75   &  0.38   &   11    &     48     &   C$^g$  & 18 & 08 & 09.72 & $-$18 & 29 & 03.8 &     45    &      48   &  EG-U    &  $\ast$   \\
B1808$-$209  &  0.99   &  0.39   &  150    &    315     &   C$^g$  & 18 & 08 & 07.47 & $-$20 & 55 & 45.1 &    290    &     315   &  EG-SR   &  $\ast$   \\
B1808$-$217  &  1.09   &  0.57   &  153    &     59     &   W      & 18 & 08 & 35.22 & $-$21 & 45 & 20.5 &     17    &      25   &  EG-D    &           \\
            &         &         &         &            &   E      &    &    & 35.36 &       &    & 18.5 &     21    &      34   &          &           \\
B1811$-$112  &  0.81   &  0.62   &    1    &    164     &   W      & 18 & 11 & 47.85 & $-$11 & 13 & 07.6 &     29    &      57   &  EG-D    &           \\
            &         &         &         &            &   E      &    &    & 49.66 &       &    & 16.9 &     45    &      89   &          &           \\
B1817$-$125  &  0.79   &  0.60   &    7    &     93     &   W      & 18 & 17 & 42.81 & $-$12 & 30 & 52.2 &     14    &      49   &  EG-D    &           \\
            &         &         &         &            &   E      &    &    & 44.96 &       &    & 45.2 &     11    &      44   &          &           \\
B1817$-$098  &  0.61   &  0.40   &   11    &    349     &   W      & 18 & 17 & 52.70 & $-$09 & 48 & 39.7 &     34    &      50   &  EG-T    &           \\
            &         &         &         &            &   C      &    &    & 52.75 &       &    & 40.6 &    110    &     121   &          &           \\
            &         &         &         &            &   E      &    &    & 52.83 &       &    & 41.0 &    118    &     169   &          &           \\
B1817$-$118  &  0.74   &  0.59   &    2    &    199     &   W      & 18 & 17 & 53.84 & $-$11 & 48 & 43.7 &     34    &     125   &  EG-D    &           \\
            &         &         &         &            &   E      &    &    & 54.04 &       &    & 43.4 &     13    &      68   &          &           \\
B1819$-$142  &  0.75   &  0.46   &  155    &    566     &   RS     & 18 & 19 & 05.39 & $-$14 & 14 & 56.3 &     24    &     566   &G-\hbox{H\,{\sc ii}} &$\ast$\\
B1819$-$131  &  0.80   &  0.59   &    9    &    185     &   W$^g$  & 18 & 19 & 21.56 & $-$13 & 11 & 14.2 &    130    &     162   &  EG-D    &           \\
            &         &         &         &            &   E      &    &    & 21.74 &       &    & 17.2 &     13    &      18   &          &           \\
B1819$-$096  &  0.76   &  0.58   &    0    &   2337     &   C$^g$  & 18 & 19 & 43.61 & $-$09 & 40 & 28.7 &   2095    &    2337   &  EG-SR   &  $\ast$   \\
B1820$-$131  &  0.80   &  0.60   &   10    &    118     &   W      & 18 & 20 & 24.62 & $-$13 & 06 & 13.4 &     21    &      42   &  EG-D    &           \\
            &         &         &         &            &   E      &    &    & 25.10 &       &    & 10.4 &     35    &      75   &          &           \\
B1822$-$149  &  0.82   &  0.59   &    7    &    119     &   N$^g$  & 18 & 22 & 43.48 & $-$14 & 57 & 40.7 &    103    &     111   &  EG-D    &           \\
            &         &         &         &            &   S      &    &    & 43.38 &       &    & 41.9 &    3.7    &     5.8   &          &           \\
B1826$-$110  &  0.82   &  0.60   &  162    &     83     &   W$^g$  & 18 & 26 & 20.60 & $-$11 & 03 & 25.7 &     58    &      66   &  EG-D    &           \\
            &         &         &         &            &   E      &    &    & 20.71 &       &    & 24.6 &    9.5    &      13   &          &           \\
B1827$-$073  &  0.79   &  0.60   &  160    &     16     &   C$^g$  & 18 & 27 & 41.15 & $-$07 & 22 & 11.7 &     13    &      16   &  EG-SR   &  $\ast$   \\
B1829$-$106  &  0.71   &  0.56   &  177    &   1198     &   N$^g$  & 18 & 29 & 34.65 & $-$10 & 37 & 26.1 &   1089    &    1148   &  EG-C+E  &  $\ast$   \\
            &         &         &         &            &   S$^g$  &    &    & 34.67 &       &    & 27.9 &     25    &      33   &          &           \\
B1830$-$089  &  0.71   &  0.62   &  168    &     81     &   N      & 18 & 30 & 35.22 & $-$08 & 57 & 31.7 &    3.7    &      28   &  EG-D    &  $\ast$   \\
            &         &         &         &            &   S$^g$  &    &    & 35.37 &       &    & 47.0 &     32    &      45   &          &           \\
B1830$-$075  &  0.72   &  0.57   &    1    &    233     &   N      & 18 & 30 & 38.65 & $-$07 & 33 & 38.9 &     18    &     168   &  EG-D    &           \\
            &         &         &         &            &   S      &    &    & 38.84 &       &    & 46.7 &     10    &      62   &          &           \\
B1832$-$094  &  0.68   &  0.63   &    0    &     63     &   W      & 18 & 32 & 13.37 & $-$09 & 27 & 35.1 &     15    &      24   &  EG-D    &           \\
            &         &         &         &            &   E      &    &    & 15.37 &       &    & 22.4 &    9.9    &      39   &          &           \\
B1832$-$044  &  0.70   &  0.63   &   20    &     58     &   W      & 18 & 32 & 31.26 & $-$04 & 24 & 41.6 &    8.7    &      11   &  EG-D    &           \\
            &         &         &         &            &   E$^g$  &    &    & 31.98 &       &    & 42.4 &     42    &      47   &          &           \\
B1832$-$113  &  0.78   &  0.63   &   16    &    118     &   N      & 18 & 32 & 32.73 & $-$11 & 18 & 23.4 &     15    &      43   &  EG-T    &           \\
            &         &         &         &            &   C      &    &    & 32.60 &       &    & 27.7 &     50    &      53   &          &           \\
            &         &         &         &            &   S      &    &    & 32.49 &       &    & 33.3 &    7.6    &      22   &          &           \\
B1834$-$044  &  0.71   &  0.62   &   27    &    109     &   W      & 18 & 34 & 25.56 & $-$04 & 26 & 54.8 &    3.3    &      26   &  EG-T    &           \\
            &         &         &         &            &   C      &    &    & 27.01 &       &    & 55.0 &    3.5    &     4.6   &          &           \\
            &         &         &         &            &   E$^g$  &    &    & 28.56 &       &    & 53.4 &     51    &      72   &          &           \\
B1834$-$018  &  0.70   &  0.63   &   27    &    220     &   S      & 18 & 34 & 42.42 & $-$01 & 53 & 01.6 &     32    &     220   &  EG-D?   &  $\ast$   \\
\end{tabular}
\end{table*}

\begin{table*}
\contcaption{}
\begin{tabular}{l rrr r l rrr rrr  rr l l }
\hline
Source & \multicolumn{3}{c}{Beam size} & S$_t$ & Cmp. & \multicolumn{3}{c}{RA(B1950)} & \multicolumn{3}{c}{Dec.(B1950)} & S$_p^c$ & S$_t^c$ & Class.&
Notes  \\
            &  maj    &  min    &   PA    &      &   & & & & & & & & & & \\
        & $^{\prime\prime}$ & $^{\prime\prime}$ & $^{\circ}$ & mJy & & h & m & s & $^{\circ}$ & $^{\prime}$ & $^{\prime\prime}$ & mJy/b & mJy & & \\
\hline
B1835$-$069A &  0.67   &  0.57   &  177    &     80     &   S$^g$  & 18 & 35 & 09.83 & $-$06 & 56 & 25.1 &     29    &      30   &  EG-D    &           \\
            &         &         &         &            &   N      &    &    & 09.80 &       &    & 14.9 &    5.2    &$\sim$50   &          &           \\
B1835$-$069B &  0.67   &  0.57   &  177    &    278     &   N      & 18 & 35 & 16.19 & $-$06 & 56 & 09.9 &     35    &      52   &  EG-D    &           \\
            &         &         &         &            &   S      &    &    & 16.24 &       &    & 11.3 &    169    &     226   &          &           \\
B1835$-$060  &  0.88   &  0.52   &   28    &     16     &   C$^g$  & 18 & 35 & 20.05 & $-$06 & 05 & 06.3 &    8.9    &      16   &  EG-SR   &  $\ast$   \\
B1835$-$063  &  0.75   &  0.56   &  170    &     62     &   W      & 18 & 35 & 23.61 & $-$06 & 21 & 48.1 &    2.7    &      24   &  EG-D    &           \\
            &         &         &         &            &   E      &    &    & 23.81 &       &    & 50.6 &    2.1    &      38   &          &           \\
B1836$-$015  &  0.84   &  0.69   &    9    &     44     &   RS     & 18 & 36 & 04.30 & $-$01 & 32 & 36.1 &    7.3    &      44   &  EG-HT?  &           \\
B1838$-$072  &  0.69   &  0.61   &   17    &     94     &   N      & 18 & 38 & 48.43 & $-$07 & 17 & 09.7 &    8.1    &      35   &  EG-D    &  $\ast$   \\
            &         &         &         &            &   S      &    &    & 48.73 &       &    & 14.7 &     14    &      55   &          &           \\
B1839$-$093  &  0.52   &  0.39   &  159    &     24     &   W      & 18 & 39 & 59.81 & $-$09 & 19 & 49.3 &     10    &      16   &  EG-T    &           \\
            &         &         &         &            &   C      &    &    & 59.94 &       &    & 48.0 &    2.7    &     4.7   &          &           \\
            &         &         &         &            &   E      &    & 40 & 00.02 &       &    & 46.8 &    1.1    &     1.7   &          &           \\
B1840$-$079  &  0.61   &  0.42   &   25    &    518     &   W      & 18 & 40 & 02.48 & $-$07 & 59 & 15.4 &     62    &     163   &  EG-D    &           \\
            &         &         &         &            &   E      &    &    & 06.92 &       &    & 05.7 &    119    &     355   &          &           \\
B1842$-$066  &  0.63   &  0.40   &  154    &     44     &   N$^g$  & 18 & 42 & 08.34 & $-$06 & 37 & 42.4 &     13    &      16   &  EG-D    &           \\
            &         &         &         &            &   S$^g$  &    &    & 08.34 &       &    & 43.1 &     26    &      28   &          &           \\
B1843$-$001  &  0.55   &  0.43   &   32    &    486     &   W      & 18 & 43 & 29.84 & $-$00 & 06 & 54.6 &     35    &      55   &  EG-T    &           \\
            &         &         &         &            &   C      &    &    & 29.89 &       &    & 53.3 &    333    &     364   &          &           \\
            &         &         &         &            &   E      &    &    & 29.95 &       &    & 52.4 &    8.7    &      51   &          &           \\
B1844$-$025  &  0.55   &  0.41   &  163    &    558     &   RS     & 18 & 44 & 23.76 & $-$02 & 31 & 11.2 &    251    &     558   &  G/EG?   &  $\ast$   \\
B1845$-$020  &  0.56   &  0.40   &  165    &    960     &   RS     & 18 & 45 & 05.66 & $-$02 & 03 & 45.9 &     17    &     960&G-\hbox{H\,{\sc ii}}/PN&$\ast$\\
B1847$-$016  &  0.70   &  0.62   &   31    &    508     &   N      & 18 & 47 & 23.48 & $-$01 & 36 & 24.6 &     37    &     142   &  EG-D    &           \\
            &         &         &         &            &   S      &    &    & 23.74 &       &    & 28.9 &    236    &     351   &          &           \\
B1848$+$039A &  0.48   &  0.43   &   26    &     27     &   N      & 18 & 48 & 06.40 & $+$03 & 55 & 31.3 &    7.5    &      22   &  EG-D    &           \\
            &         &         &         &            &   S      &    &    & 06.25 &       &    & 28.9 &    2.4    &     3.7   &          &           \\   
B1848$+$039B &  0.48   &  0.43   &   26    &     43     &   N      & 18 & 48 & 10.97 & $+$03 & 54 & 02.5 &     12    &      20   &          &           \\
            &         &         &         &            &   C      &    &    & 10.93 &       &    & 01.2 &    1.8    &     2.6   &          &           \\
            &         &         &         &            &   S      &    &    & 10.86 &       & 53 & 59.1 &     12    &      19   &          &           \\
B1848$+$029  &  0.56   &  0.43   &   36    &    173     &   W$^g$  & 18 & 48 & 44.11 & $+$02 & 56 & 02.2 &     58    &      63   &  EG-D    &           \\
            &         &         &         &            &   E$^g$  &    &    & 44.14 &       &    & 02.0 &    102    &     109   &          &           \\
B1849$+$005  &  0.54   &  0.42   &  161    &    622     &   N      & 18 & 49 & 13.57 & $+$00 & 31 & 52.8 &    487    &     549   &  EG-C+E  &  $\ast$   \\
            &         &         &         &            &   S      &    &    & 13.43 &       &    & 49.1 &     19    &      52   &          &           \\
B1850$+$011  &  0.57   &  0.50   &  152    &   1107     &   RS     & 18 & 50 & 46.15 & $+$01 & 11 & 12.1 &     49    &    1107   &G-\hbox{H\,{\sc ii}}&$\ast$\\
B1851$+$031A &  0.63   &  0.62   &   26    &     60     &   N      & 18 & 51 & 06.04 & $+$03 & 10 & 29.3 &     12    &      24   &  EG-T    &           \\
            &         &         &         &            &   C      &    &    & 06.10 &       &    & 28.2 &    8.2    &      14   &          &           \\
            &         &         &         &            &   S      &    &    & 06.17 &       &    & 25.8 &    6.0    &      13   &          &           \\
B1851$+$031B &  0.63   &  0.62   &   26    &     60     &   W      & 18 & 51 & 10.23 & $+$03 & 11 & 54.9 &     15    &      35   &  EG-T    &  $\ast$   \\
            &         &         &         &            &   C      &    &    & 10.70 &       &    & 51.3 &    2.1    &           &          &           \\
            &         &         &         &            &   E      &    &    & 11.40 &       &    & 48.0 &    4.6    &      23   &          &           \\
B1853$+$028  &  0.56   &  0.44   &   37    &    107     &   N      & 18 & 53 & 21.06 & $+$02 & 52 & 55.1 &     19    &      38   &  EG-D    &           \\
            &         &         &         &            &   S      &    &    & 21.79 &       &    & 34.1 &     16    &      69   &          &           \\
B1855$+$031  &  0.68   &  0.61   &  176    &    842     &   C      & 18 & 55 & 32.16 & $+$03 & 09 & 10.1 &    762    &     842   &  EG-C+E  &           \\
B1857$-$000  &  0.56   &  0.42   &  156    &    197     &   N$^g$  & 18 & 57 & 43.81 & $-$00 & 00 & 19.9 &    148    &     175   &  EG-C+E  &  $\ast$   \\
            &         &         &         &            &   S      &    &    & 43.79 &       &    & 22.9 &    1.9    &      12   &          &           \\
B1901$+$114  &  0.46   &  0.40   &  172    &     30     &   N      & 19 & 00 & 58.95 & $+$11 & 25 & 25.8 &    8.5    &      26   &  EG-D?   &           \\
            &         &         &         &            &   S      &    &    & 59.15 &       &    & 22.1 &    1.6    &     2.9   &          &           \\
B1901$+$058  &  0.67   &  0.60   &    0    &    155     &   N$^g$  & 19 & 01 & 16.07 & $+$05 & 48 & 27.3 &     64    &      72   &  EG-D    &           \\
            &         &         &         &            &   S$^g$  &    &    & 16.14 &       &    & 26.2 &     79    &      83   &          &           \\
B1910$+$161  &  0.65   &  0.61   &    4    &     54     &   W      & 19 & 10 & 04.45 & $+$16 & 11 & 19.2 &    3.1    &      13   &  EG-T    &           \\
            &         &         &         &            &   C      &    &    & 04.52 &       &    & 20.1 &     12    &      14   &          &           \\
            &         &         &         &            &   E      &    &    & 04.68 &       &    & 22.9 &     12    &      12   &          &           \\
B1913$+$115  &  0.46   &  0.40   &  171    &     35     &   W      & 19 & 13 & 19.36 & $+$11 & 33 & 02.1 &    5.0    &      10   &  EG-D    &           \\
            &         &         &         &            &   E      &    &    & 19.55 &       &    & 02.3 &    5.5    &      20   &          &           \\
B1921$+$096  &  0.66   &  0.60   &    6    &    101     &   N      & 19 & 21 & 31.55 & $+$09 & 38 & 42.9 &    6.1    &      35   &  EG-D    &           \\
            &         &         &         &            &   S      &    &    & 32.74 &       &    & 21.3 &     19    &      66   &          &           \\
B1922$+$138  &  0.43   &  0.39   &    0    &    253     &   W      & 19 & 22 & 58.17 & $+$13 & 53 & 25.1 &     62    &     145   &  EG-D    &           \\
            &         &         &         &            &   E      &    & 23 & 00.07 &       &    & 10.6 &     22    &     108   &          &           \\
B1932$+$204  &  0.66   &  0.60   &   15    &    335     &   C$^g$  & 19 & 32 & 59.56 & $+$20 & 25 & 14.1 &    337    &     335   &  EG-U    &           \\
B1933$+$167  &  0.77   &  0.60   &   83    &    223     &   C      & 19 & 33 & 30.18 & $+$16 & 42 & 09.1 &    185    &     223   &  EG-SR   &  $\ast$   \\
B1938$+$229  &  0.41   &  0.39   &   26    &     62     &   RS     & 19 & 38 & 42.05 & $+$22 & 59 & 03.1 &     10    &      62   &  G/EG?   &  $\ast$   \\
\end{tabular}
\end{table*}

Following Sramek et al. (1992), any young SNR candidate would be characterised by a shell-like structure and
a non-thermal radio spectrum. The non-thermal spectrum would distinguish a young SNR from
a Galactic \hbox{H\,{\sc ii}} region or planetary nebula. 
The images of the extended sources with shell or partial shell-like
structures (B1819$-$142, B1845$-$020 and B1850$+$011) and which, from a combination of our measurements and
those at other frequencies in the literature, have been shown to have thermal spectra, are presented in Figs. 1 and 2. 
Those exhibiting any resemblance to shell- or ring-like structures with non-thermal spectra (B1801$-$203, B1802$-$196 and B1938$+$229)
are shown in Figs. 3 and 4. The VLA images of the remaining well-resolved sources are shown in Fig. 5, and a 
MERLIN 5 GHz image of B1857$-$000 is presented in Fig. 6. These latter sources are all thought to be of
extragalactic origin.

\begin{figure}
\vbox{
  \psfig{file=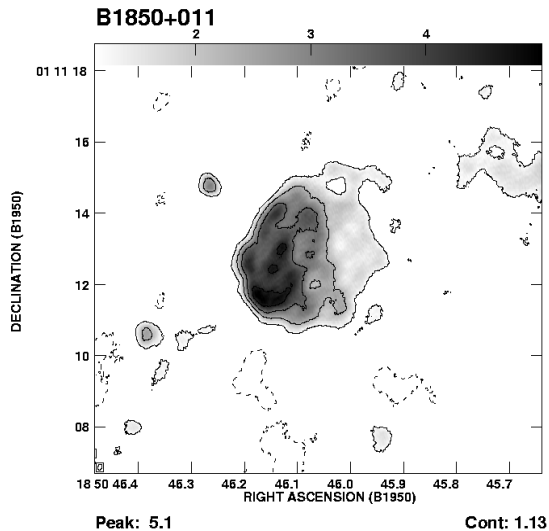,height=3.0in}
\caption[]{MERLIN$+$VLA image at $\lambda$6 cm of the ultracompact \hbox{H\,{\sc ii}} region B1850$+$011
(G34.26$+$0.15) with a resolution of 171$\times$116 mas along a PA of $-$20.5$^\circ$.
The contours are $-$1, 1, 2, 3, 4 \ldots times the first contour.
The peak brightness and the level of the first contour in units of mJy/beam are shown below the image.}
}
\end{figure}

\subsection{The Galactic thermal sources}
\noindent
{\bf B1819$-$142, G16.94$-$0.07}: The source has a counterpart in the IRAS Point Source Catalogue (PSC), and its 
                  IRAS colours satisfy
                  the Wood \& Churchwell (1989) criterion for classifying it as an
                  ultracompact(UC) \hbox{H\,{\sc ii}} region.
                  It has an inverted spectrum between 1400 and 5000 MHz (Garwood et al.
                  1988; Z90; B94). CS(2$-$1)
                  emission has been detected from this UC \hbox{H\,{\sc ii}} 
                  region (Bronfman, Nyman \& May 1996).

\noindent
{\bf B1845$-$020, G30.71$-$0.08}: There is no counterpart in the IRAS PSC. From the flux density measured from
                 this data and values quoted by Garwood et al. 1988; Z90 and B94, the source is seen to have 
                 an inverted spectrum between 1400 and 5000 MHz                  
                 and could be either an UC \hbox{H\,{\sc ii}} region or, more probably, a planetary nebula. Its 
                 structure appears very similar to that of the young planetary nebula BD +303639
                 (Bryce et al. 1997).  
                  
\noindent
{\bf B1850$+$011, G34.26$+$0.15}:  This source, which has been extensively studied, is an
                 UC \hbox{H\,{\sc ii}} region with a cometary morphology, though with two very compact
                 components approximately 2 arcsec to the east of the main extended radio emission.
                 Using the nomenclature of Gaume, Fey \& Claussen (1994), the compact components are referred to as
                 A and B, and the extended emission as component C.
                 Since the first high-resolution radio continuum image was published by Reid \& Ho (1985)
                 at 18cm, images of the source over a range of wavelenghts from 18cm to 1.3cm have been
                 made. Spectral lines due to a wide range of molecules including OH, H$_2$O, NH$_3$, CO
                 and SiO as well as mm, submm and near, mid and far infrared continuum observations have
                 led to a considerable understanding of the associated molecular clouds, with the
                 hottest, densest molecular region  being located between the cometary ionisation front
                 and the two compact components to the east of the extended emission (G\'{o}mez  et al.
                 2000; Campbell et al. 2004). H76$\alpha$
                 (Garay, Rodr\'{i}guez \& van Gorkom 1986) and H93$\alpha$ (Gaume et al. 1994) radio recombination lines have
                 shown a velocity gradient from north to south across the extended component. This has
                 been interpreted by Garay et al. (1986) as a rotation of a molecular disk round a central 
                 object, probably a cluster of stars.  Gaume et al. (1994) consider it to arise from
                 the momentum fluxes of stellar winds from the compact
                 components impinging upon the cometary ionized region at differing position angles.
                 The spectral indices of the two compact sources, A and B, are 
                 consistent with an ionization bounded stellar wind (Gaume et al. 1994;
                 Wood \& Churchwell 1989). The VLA+MERLIN image clearly resolves them, but an increased
                 signal to noise is required to determine definitive sizes. The southern peak in the head
                 of component C is clearly resolved into an arc of emission pointing towards the south.

\subsection{The Galactic non-thermal source}

\noindent
{\bf B1749$-$281, G1.54$-$0.96}: The measured position of the source is within $\sim$0.7  arcsec of that of a
                 known pulsar (Taylor, Manchester \& Lyne 1993; Han \& Tian 1999). It would seem that
                 the continuum source seen by us represents a mean flux density level for that of the
                 pulsar. However, the pulsar has a very steep spectrum ($\alpha\sim$2.8) using the
                 listed flux densities of 1150 and 35 mJy at 400 and 1400 MHz respectively (Taylor et al. 1993).
                 At 1.4 GHz, the estimated flux density from the NVSS is 41.6$\pm$1.3 mJy 
                 (Han \& Tian 1999; Kaplan et al. 1998). Using this value, the expected flux density at
                 5 GHz is $\sim$1.5 mJy. The 
                 measured flux density of the source is 6.1 mJy at 4835 MHz and 1.6 mJy at 4885 MHz. 
                 With a dispersion measure of 51$\pm$14 cm$^{-3}$ pc (Taylor et al. 1993) the pulsar would
                 probably be close to its transition frequency between strong and weak scintillation
                 at 5 GHz. If that is the case, very large modulations can occur since both
                 diffractive and refractive scintillation contribute to the modulation with almost equal 
                 strength (Kramer, private communication). 
                 The fact that Kaplan, Cordes \& Condon (2000) did not detect any source at this position 
                 in their continuum survey of pulsar positions at 5GHz with an estimated flux density of
                 0.0$\pm$0.08 mJy, is in agreement with this.

\begin{figure}
  \vbox{
  \psfig{file=F1801-203G.PS,width=2.1in}
  \psfig{file=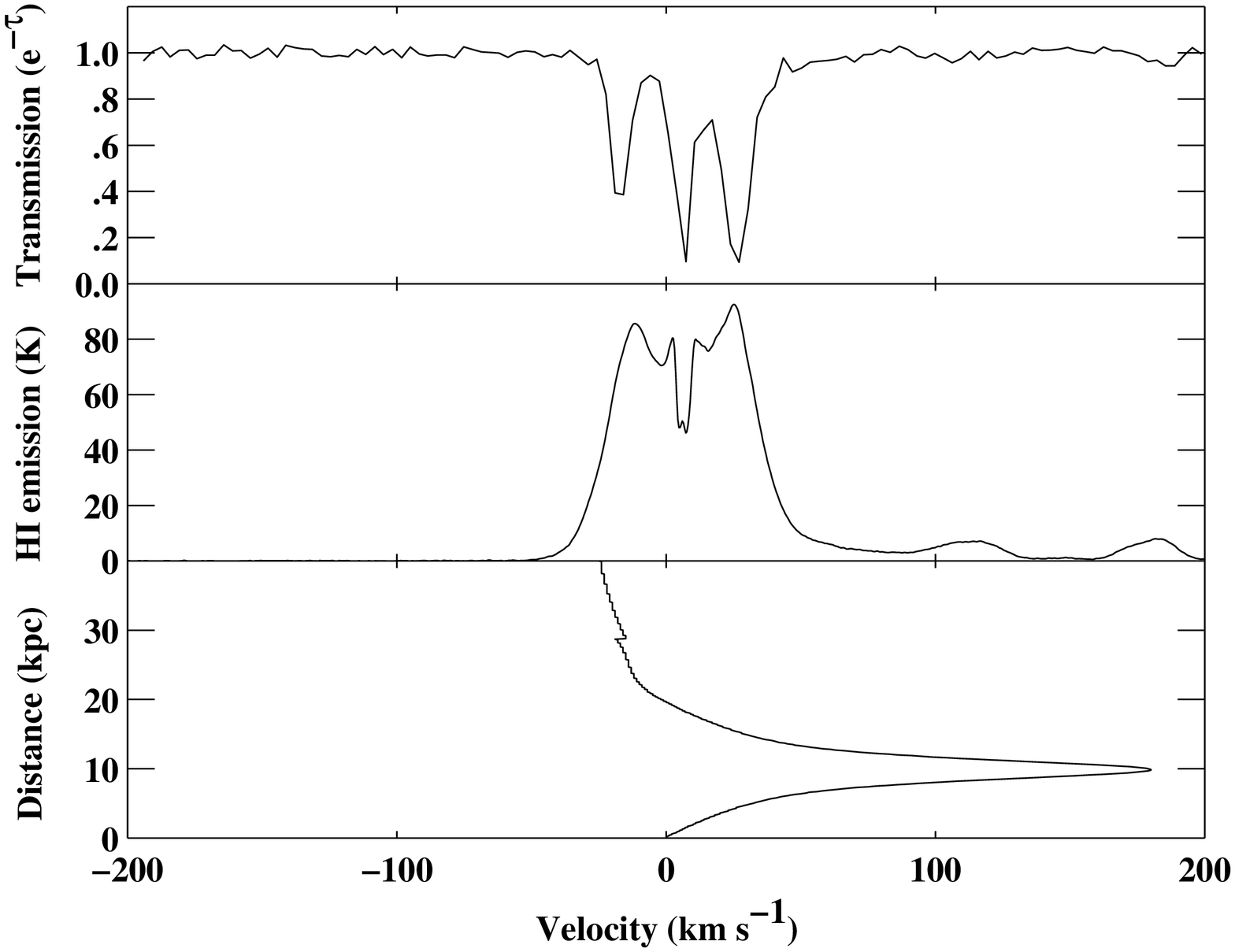,width=2.52in,angle=-90}
  \psfig{file=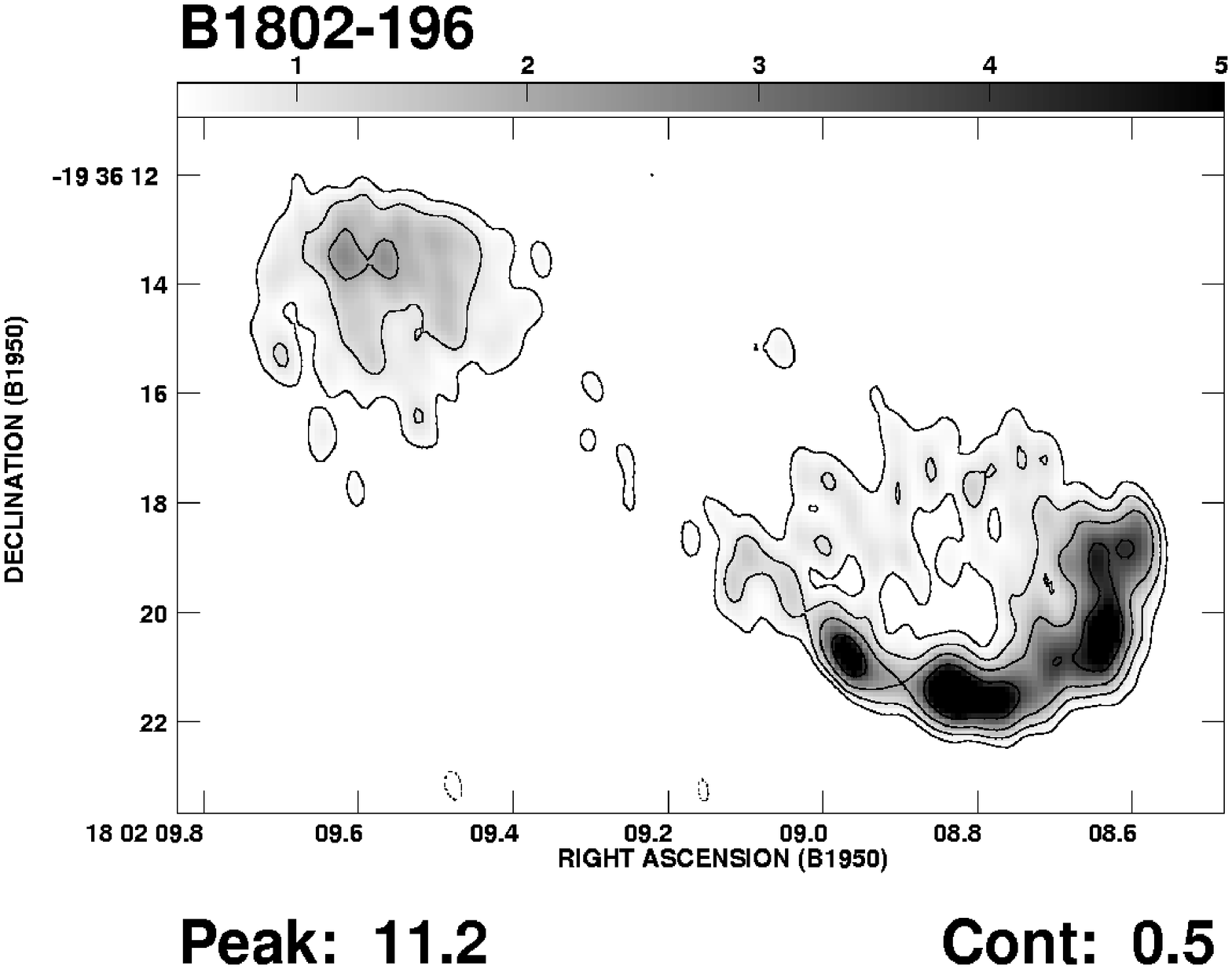,width=2.1in,angle=-90}
  \psfig{file=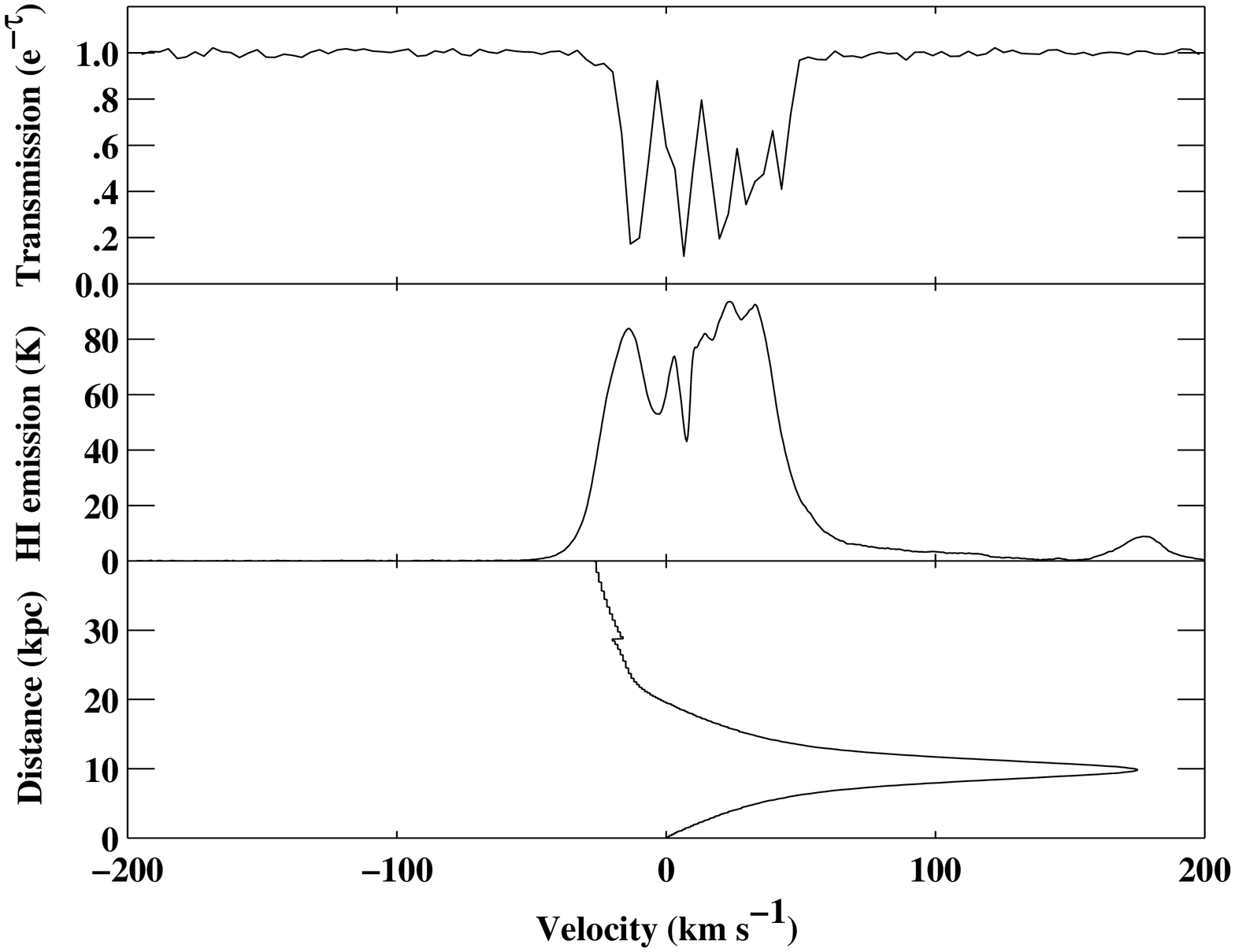,width=2.52in,angle=-90}
  }
\caption[]{VLA continuum image of B1801$-$203 and B1802$-$196
with angular resolutions of $\sim$0.4 and 0.5 arcsec respectively (see Table 1). 
The contours are $-$1, 1, 2, 4 \ldots times
the first contour in units of mJy/beam shown below each image. The peak brightness 
is also in units of mJy/beam.
The corresponding GMRT \hbox{H\,{\sc i}} absorption spectra obtained
with resolutions of 5.4$\times$4.1 arcsec along PA=104$^\circ$ 
and 17.1$\times$12.6 arcsec along PA=102$^\circ$ respectively,
the \hbox{H\,{\sc i}} emission spectrum from the Leiden/Dwingeloo survey
and the distance-velocity profiles in the direction of 
the sources are shown below the continuum images. The velocities
in Figs. 3 and 4 are relative to the local standard of rest.
}
\end{figure}

\begin{figure}
\vbox{
\hbox{
  \psfig{file=F1938+229_P2.PS,width=3.0in,angle=-90}
     }
\hbox{
  \psfig{file=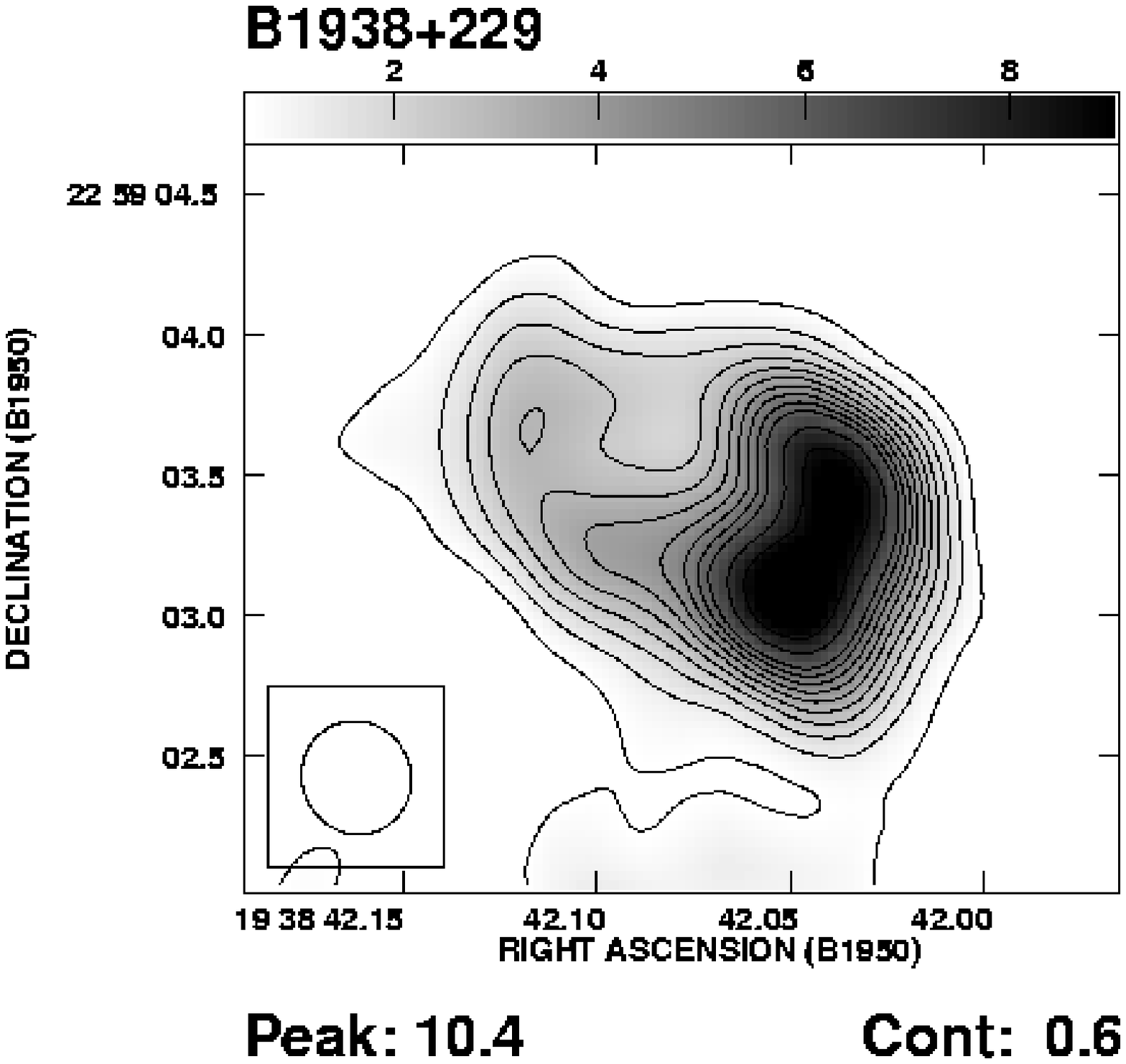,width=1.6in}
  \psfig{file=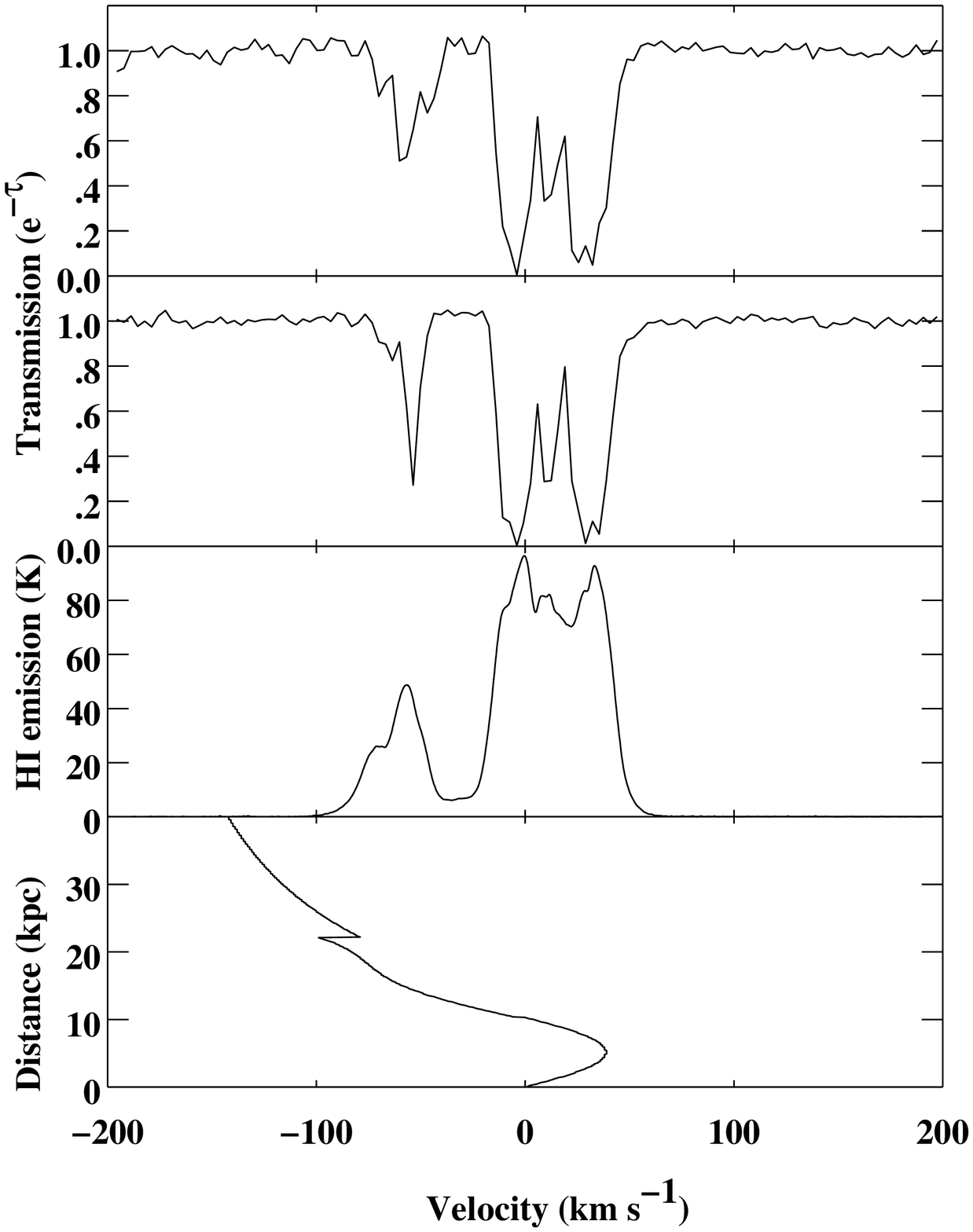,width=1.7in}
     }
}
\caption[]{Top: The GMRT continuum image of 1938$+$229 at a frequency of 
1420  MHz and a resolution of 9.5$\times$5.0 arcsec along PA=114$^\circ$.
 Bottom left: The VLA image at a resolution of $\sim$0.4 arcsec (see Table 1).  
For both images, the contour 
levels are -1, 1, 2, 4,  \ldots mJy/beam times the
first contour shown below each image. The peak brightness is also
in units of mJy/beam. Bottom right: 
The GMRT \hbox{H\,{\sc i}} absorption spectra towards the two lobes,
the \hbox{H\,{\sc i}} emission spectrum from the Leiden/Dwingeloo survey
and the distance-velocity profile in the direction of
the source. 
}
\end{figure}

\subsection{Candidate shell structures}

\noindent
{\bf B1801$-$203, G9.55$+$0.69}: Having a spectral index of $\sim$1 between 365 and 4835 MHz (Douglas
                  et al. 1996; Z90), the present observations
                  reveal it to be a highly asymmetric double (Fig. 3). The western component 
                  has a shell-like 
                  structure, reminiscent of a  supernova remnant candidate, although the overall
                  structure resembles an asymmetric double-lobed extragalactic source. 
                  The GMRT \hbox{H\,{\sc i}} absorption spectrum (Fig. 3) has a most negative
                  absorption feature at $\sim$$-$15 km s$^{-1}$, and continues to about $\sim$$-$25 km s$^{-1}$. 
                  A comparison of the absorption spectrum with the emission spectrum from the Leiden/Dwingeloo
                  survey (Hartmann \& Burton 1996; Hartmann et al. 1996) and with the distance-velocity plot for the
                  Galactic longitude of the source (Burton 1988; Dickey \& Lockman 1990), 
                  clearly indicates that the source is extragalactic.  The column density along
                  the line of sight to the source has been estimated to be 8.9$\times$10$^{21}$ cm$^{-2}$,
                  assuming a spin temperature of 100 K.  
                  Rings or shell-like structures as seen in this source are known to occur in the lobes 
                  of a few extragalactic  radio sources, such as Her A, 3C310, 
                  MG0248+0641 (Conner et al. 1998 and references therein) and 3C459 
                  (Thomasson et al. 2003). Such structures could be caused by gravitational lensing
                  or the formation of 'bubbles' as a result of flow instabilities. 

\noindent
{\bf B1802$-$196, G10.30$+$0.88}: The spectral index of the whole
                  source is 1.0 between 365 and 4835 MHz. A lower-resolution image has
                  been published by Sramek et al. (1992). Although the structure of the
                  western lobe resembles a partial shell-like structure (Fig. 3), the GMRT
                  \hbox{H\,{\sc i}} absorption spectrum shows that the maximum absorption is at 
                  $\sim$$-$14 km s$^{-1}$ and continues to about 
                  $\sim$$-$27 km s$^{-1}$, indicating that it too is of extragalactic origin
                  (Fig. 3). The 
                  column density along the line of sight to the source has been estimated to be 
                  10$^{22}$ cm$^{-2}$.  If the high-brightness ridge of emission is interpreted
                  to be a curved radio jet, it is a one-sided jet without a detected radio
                  core, which is uncommon.

\noindent
{\bf B1938$+$229, G58.97$+$0.22}: The VLA image of B1938$+$229 (Fig. 4 lower left) shows a
                  shell-like structure.  A tapered image of the field  with a resolution of
                  $\sim$0.62$\times$0.57 arcsec$^2$ also revealed a weak component approximately 
                  72 arcsec to the south east
                  at RA 19$^h$ 38$^m$ 47.$^s$06 Dec $+$22$^\circ$ 58$^{\prime}$ 43.$^{\prime\prime}$5
                  with peak and total flux densities of 14 and 53 mJy respectively.  
                  It was not clear if the two components were related, but GMRT continuum and \hbox{H\,{\sc i}}
                  observations show that they are and that the whole source
                  is a double-lobed extragalactic one (Fig. 4, upper and lower right). 
                  From these observations and the data of Garwood et al. (1988), Z90 and Taylor et al. (1996) 
                  the spectral index of the source is $\sim$0.9 between 327 and 5000 MHz. 
                  The maximum absorption in the western lobe occurs at about $\sim$$-$54 km s$^{-1}$ and continues
                  to $\sim$$-$70 km s$^{-1}$, while for the eastern lobe the corresponding values are 
                  $\sim$$-$60 km s$^{-1}$ and $\sim$$-$70 km s$^{-1}$ respectively. 
                  The differences in the \hbox{H\,{\sc i}} absorption line profiles against the two lobes
                  can be attributed to variations in the optical depth, $\tau$, as the source is seen through the 
                  lower edge of the warped Galactic \hbox{H\,{\sc i}} disk (Hartmann \& Burton 1996; Hartmann et al. 1996). 
                  The column densities 
                  in the directions of the two lobes are $\sim$2$\times$10$^{22}$ cm$^{-2}$.

\begin{figure*}
\psfig{file=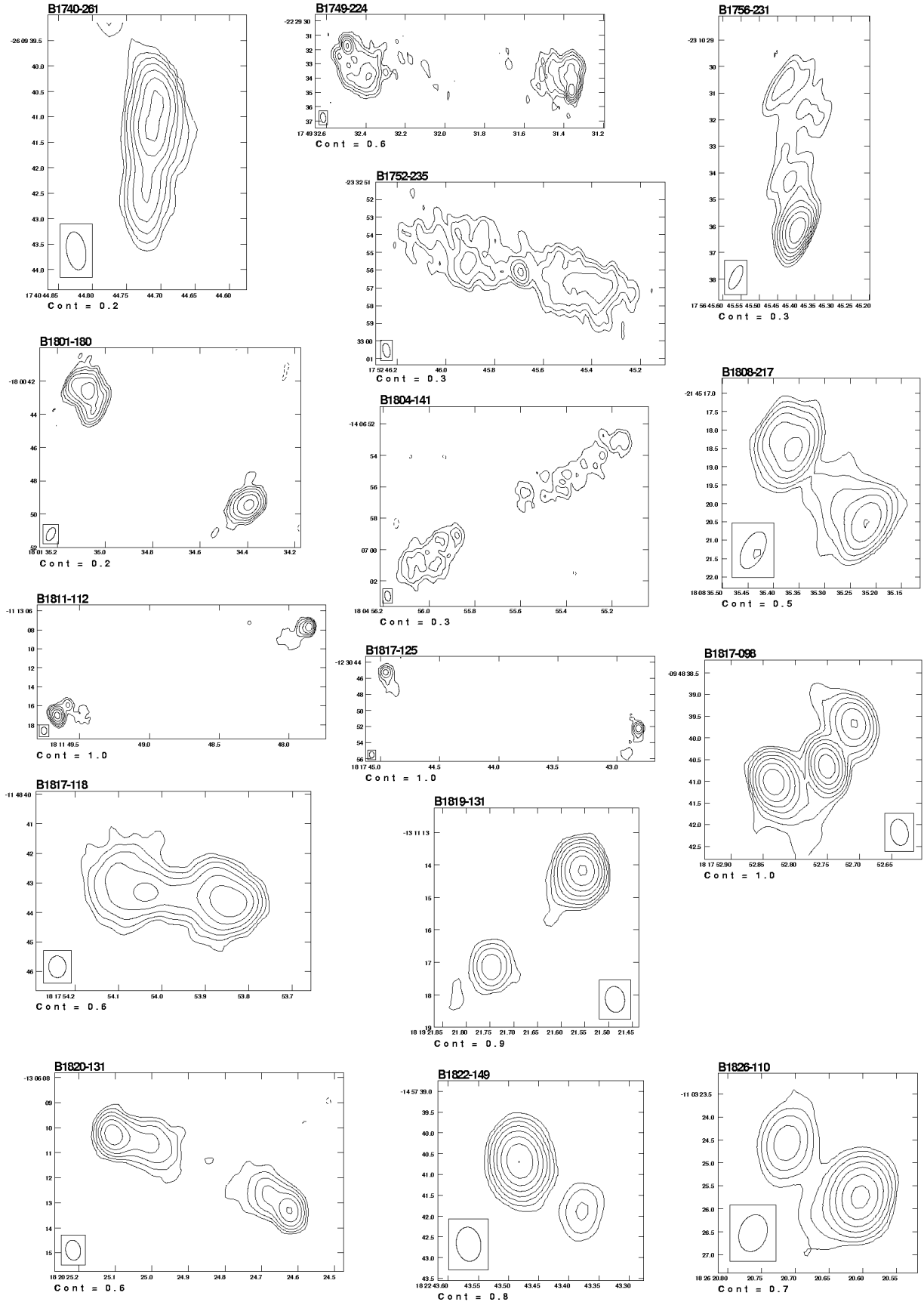,width=6.5in}
\caption[]{VLA images of the sources at $\lambda$6 cm. The contours are $-$1, 1, 2, 4 \ldots times
the first contour in units of mJy/beam shown below each image, except for B1835$-$063 and B1913$+$115 where the contours are
$-$1, 1, 2, 3, 4 \ldots times the first contour. The positions are in B1950 co-ordinates.}
\end{figure*}

\begin{figure*}
\psfig{file=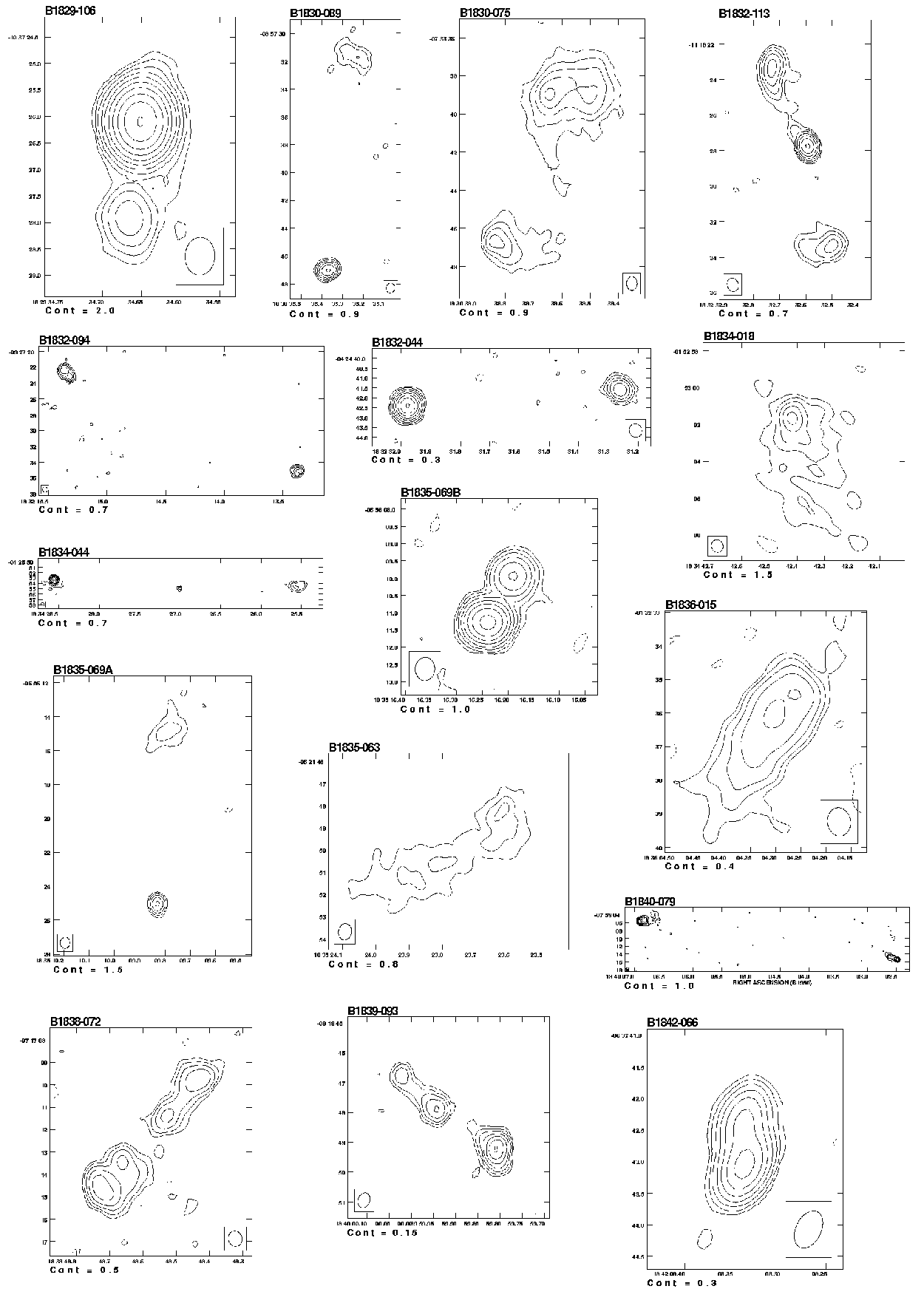,width=6.5in}
\contcaption{}
\end{figure*}

\begin{figure*}
\psfig{file=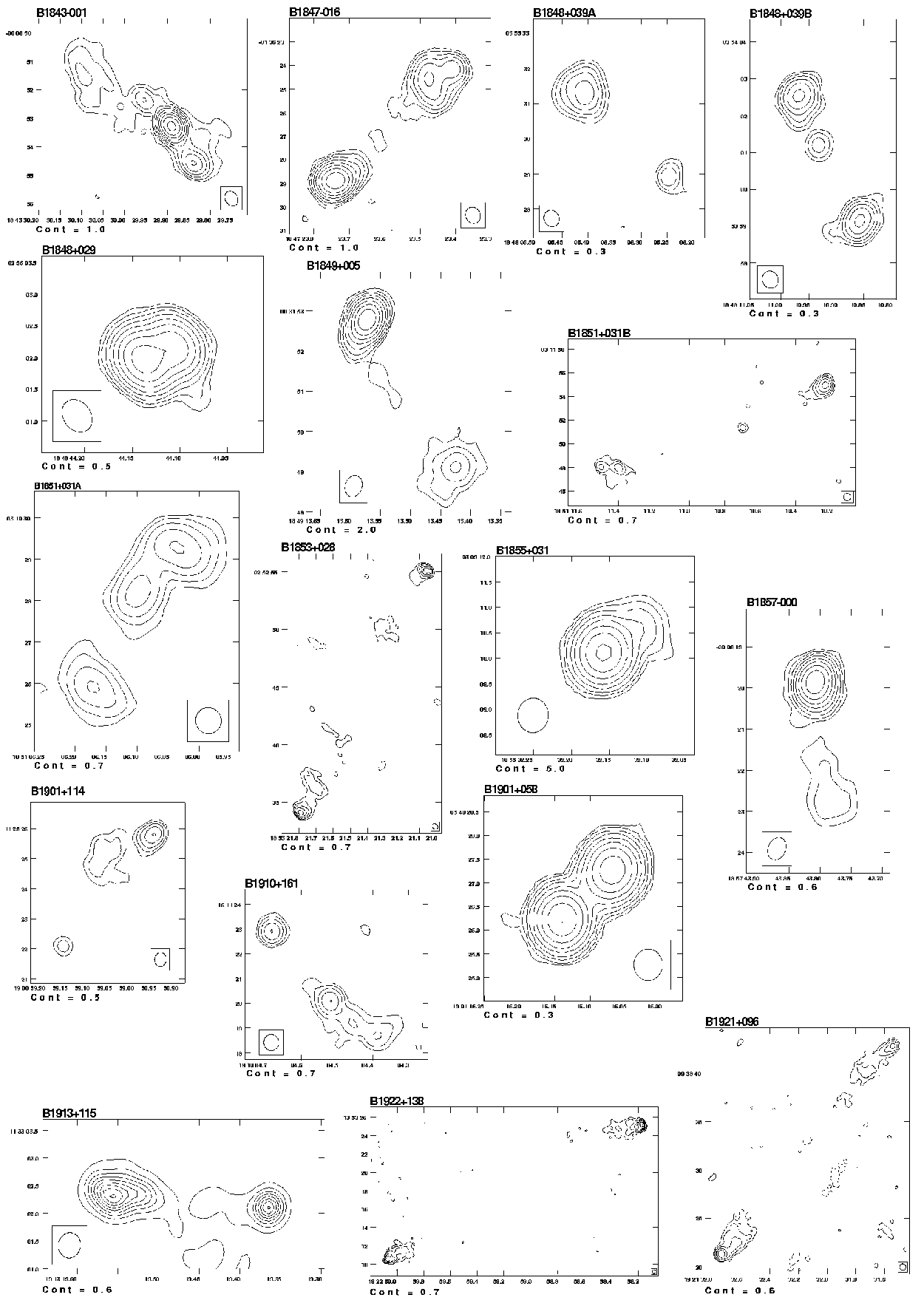,width=6.5in}
\contcaption{}
\end{figure*}

\subsection{The remaining sources}

\noindent
{\bf B1748$-$253}: The source is resolved with an angular size of $\sim$0.2 arcsec. It has been 
                  observed earlier by Lazio \& Cordes (1998).
 
\noindent
{\bf B1754$-$226}: There is a second source in the
                  field at RA 17$^h$ 54$^m$ 27.$^s$16, Dec $-$22$^\circ$ 41$^{\prime}$ 
                  14.$^{\prime\prime}$7 with a flux density of 48 mJy.

\noindent
{\bf B1808$-$184}: Its flux density measured by B94 is only 29 mJy, 
                  suggesting that it is strongly variable.

\noindent
{\bf B1808$-$209}: It is resolved with an angular size of $\sim$0.1 arcsec.

\noindent
{\bf B1819$-$096}: The source is resolved and appears elongated with an angular size of 
                  0.3$\times$0.1 arcsec$^2$ along a PA of 36$^\circ$.

\noindent
{\bf B1827$-$073}: The source is resolved with an angular size of $\sim$0.3 arcsec.

\noindent
{\bf B1829$-$106}: The radio spectrum is flat ($\alpha_{365}^{4835} \sim 0.1$) and the
                  structure is similar to that of a core-dominated, one-sided radio source.

\noindent
{\bf B1830$-$089}: The southern component is resolved, has a steep spectrum and is likely to
                  be a compact hotspot.

\noindent
{\bf B1834$-$018}: It is likely to be the hotspot of a highly asymmetric double-lobed source,
                  reminiscent of B0500$+$630 (Saikia et al. 1996).

\noindent
{\bf B1835$-$060}: The source is resolved with an angular size of $\sim$0.6 arcsec.

\noindent
{\bf B1838$-$072}: There is an unrelated source with peak and integrated flux densities of 
                  13 and 21 mJy respectively at RA 18$^h$ 38$^m$ 50.$^s$31 Dec $-$07$^\circ$ 
                  15$^{\prime}$ 20.$^{\prime\prime}$9.

\noindent
{\bf B1844$-$025}: The source is somewhat resolved with an estimated 
                  angular size of $\sim$0.7$\times$0.4
                  arcsec$^2$ along PA of 9$^\circ$. It has a flat spectrum between 365
                  and 5000 MHz (Garwood et al. 1988; Z90; B94; Douglas et al. 1996; present paper), and is
                  $\sim$20$^{\prime\prime}$ from a region of H$_2$O maser emission
                  (Valdettaro et al.  2001). There is a source in the IRAS PSC $\sim$30
                  arcsec from the radio source, which could be associated with it. If so, it
                  could be an UC \hbox{H\,{\sc ii}} region.  However, a background 
                  flat-spectrum source heavily broadened by ISS cannot be ruled out. It is worthy
                  of further investigation.

\noindent
{\bf B1849$+$005}: The radio spectrum flattens towards higher frequencies (Z90, B94, Douglas et al. 1996), 
                  suggesting it to be a core-dominated source.                 

\noindent
{\bf B1851$+$031B}: There is a weak source with peak and integrated flux densities of 
                   $\sim$3 and 5 mJy respectively at RA 18$^h$ 51$^m$ 08.$^s$03 Dec 03$^\circ$
                   10$^{\prime}$ 33.$^{\prime\prime}$8.

\noindent
{\bf B1857$-$000}: The VLA image showed the source to have a one-sided structure with a 
                  dominant core which appeared resolved with an angular size of 
                  0.26$\times$0.13 arcsec$^2$ along a PA of 25$^\circ$. The source was observed
                  with MERLIN to investigate whether the resolved size might be due to 
                  scattering or to a prominent jet close to the nucleus. The MERLIN image at
                  5 GHz (Fig. 6) shows an unresolved core and extended jet-like emission
                  towards the south-west.
                 
\noindent
{\bf B1933$+$167}: The source is resolved with an estimated angular size of $\sim$0.3 arcsec. 

\begin{figure}
\vbox{
  \psfig{file=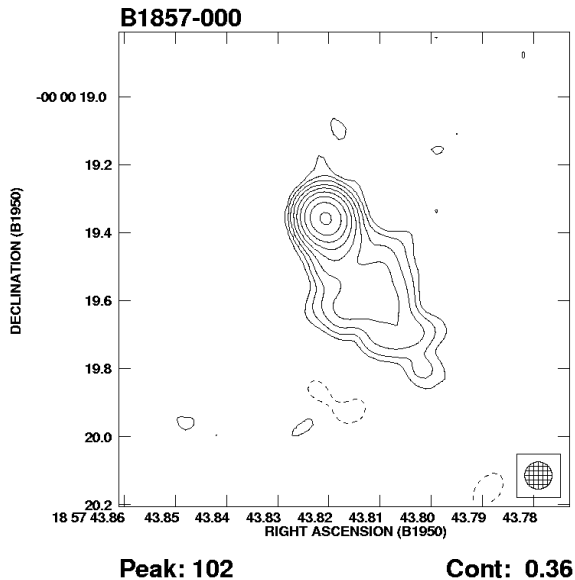,width=3.0in}
\caption[]{MERLIN image of B1857$-$000 at $\lambda$6 cm with an angular resolution of 80 mas.
The contours are $-$1, 1, 2, 4 \ldots times the first contour shown below the image.
The peak brightness and the level of the first contour are in units of mJy/beam.}
}
\end{figure}

\section{Concluding remarks}
Subarcsec-resolution observations of a sample of low galactic latitude sources have revealed shells in two
Galactic thermal radio sources, G16.94$-$0.07 and G30.71$-$0.08, and shell-like like 
structures in three non-thermal radio sources, which are likely to be extragalactic radio sources. VLA
and MERLIN data have been combined to produce a high resolution image of the UC \hbox{H\,{\sc ii}} region,
which reveals an arc-like structure  pointing towards the south. 
No good young supernova remnant candidates have been found, which is consistent with earlier 
searches (e.g. Green 1989,  Sramek et al. 1992 and references therein). 
The observations have shown that ten of the sources are single and compact, 
eight of which are slightly resolved at 5 GHz, 
marked SR in Table 1, while the remaining two are unresolved. The resolved sources may have been 
broadened by ISS, although further observations of these sources at different wavelenghts are 
required to establish the effects of ISS. One of the slighly resolved sources, B1857$-$000, was observed
with MERLIN at 5 GHz and found to have a core-jet structure.

The radio source B1749$-$281 is believed to be a pulsar which has not been detected earlier in
continuum observations at 5 GHz (Kaplan et al. 2000). The difference in the measured 
flux densities in the two 5 GHz bands, and also its earlier non-detection, is 
almost certainly due to a combination of refractive and diffractive scintillations. 

A number of the extragalactic double-lobed radio sources are highly asymmetric
in their brightness ratios. Six of them have a brightness ratio $\gapp$10 and a further nine
have a ratio $\gapp$5. These are all lobe-dominated radio sources. Assuming an angle of 
$\sim$60$^\circ$ to the line-of-sight and a typical hot-spot speed of $\sim$0.2c, the expected
ratio is only $\sim$2, suggesting that many of these sources which are of small angular size
are intrinsically asymmetric. 

\section*{Acknowledgments}
We thank Dave Green for his helpful comments on the manuscript.
The National Radio Astronomy Observatory is a facility of the National Science Foundation
operated under co-operative agreement by Associated Universities, Inc. MERLIN is a UK
National Facility operated by the University of Manchester on behalf of PPARC. 
The GMRT is a national facility operated by the National Centre for Radio Astrophysics
of the Tata Institute of Fundamental Research.  This research has 
made use of the SIMBAD database, operated at CDS, Strasbourg, France and the  
NASA/IPAC Extragalactic Database (NED), which is operated by the Jet Propulsion Laboratory,
California Institute of Technology under contract with the National Aeronautics and Space
Administration. DJS thanks the Director, Jodrell Bank Observatory for hospitality, where
most of the work was done.

{}

\end{document}